\documentstyle[11pt,newpasp_w,twoside]{article}
\def\edcomment#1{\iffalse\marginpar{\raggedright\sl#1\/}\else\relax\fi}
\reversemarginpar

\begin{document}

\title{The interplay of observational errors through numerical
simulations using the $1/V_{\rm max}$ method for magnitude and
proper-motion samples of local disk white dwarfs.}

\author{Ren\'e A. M\'endez} \affil{Cerro Tololo Inter-American
 Observatory, Casilla 603, La Serena, Chile. \\ E-mail:
 rmendez@noao.ed}

\author{Mar\'{\i}a Teresa Ruiz} \affil{Departamento de
Astronom\'{\i}a, Universidad de Chile, Casilla 36-D, Santiago,
Chile. \\ E-mail: mtruiz@das.uchile.cl}
\begin{abstract}
%
We examine the faint-end ($M_V > +14$) behavior of the disk white
dwarf luminosity function using the $1/V_{\rm max}$ method, but, for
the first time, fully including the effects of realistic observational
errors in the derived luminosity function.
We find that observational errors, mostly in the bolometric
corrections and trigonometric parallaxes, play a major role in
obliterating (real or artificial) small scale fluctuations in the
luminosity function. A better estimator of the true luminosity
function seems to be the median over simulations, rather than the
mean. When using the latter, an age of 10~Gyr or older can not be
ruled out from the sample of Leggett, Ruiz, and Bergeron (1998).
\end{abstract}
\section{Introduction}
The classical method for determining the luminosity function (LF
hereafter) of magnitude- and proper-motion selected samples is that
proposed by Schmidt (1975). This method, called the $1/V_{\rm max}$,
stems from a generalization of a method proposed earlier by Schmidt
(1968) for magnitude-limited samples. The method assumes that the LF
does not change (``evolves'') as a function of distance from the
observer, and that the sample is homogeneously distributed in
space. The $1/V_{\rm max}$ method computes the LF by ``weighting'' the
contribution of each observed point by the equivalent volume where
that particular object could have been observed under the
pre-specified survey constraints. Several modifications have been
proposed to the method in the case of magnitude-limited samples (Davis
and Huchra 1982, Eales 1993), most notably that to be able to combine
different samples coherently (Avni and Bahcall 1980). However, the
basic scheme to determine the LF of magnitude {\it and} proper-motion
selected samples has remained unchanged, and few and limited numerical
simulations have been carried-out to explore the robustness and
possible biases that the original method might have when dealing with
complete but small, kinematically selected samples (Wood and Oswalt
1998).

Additionally, prompted by investigations of the LF of galaxies where,
both, galaxy evolution and clustering (inhomogeneity) plays an
important role, several new methods have been proposed to determine
the LF of magnitude- and redshift-limited samples. These methods have
been designed so as to be less sensitive than the $1/V_{\rm max}$ to
galaxy evolution and clustering. Some of these methods are parametric,
some are not. For a review of the different methods, and their
relative merit, the reader is referred to Willmer
(1997). Unfortunately, these methods are difficult to generalize to
proper-motion selected samples as they imply the simultaneous
integration of the LF and the projected tangential velocity
distribution function, and thus they are extremely model-dependent,
unlike the $1/V_{\rm max}$ which only uses kinematic information,
without regard to the underlying velocity distribution. For example,
the Step-Wise maximum-likelihood method (Efstathiou et al. 1988),
which is one of the most robust {\it non-parametric} methods proposed
so far (Willmer 1997) is difficult to generalize to proper-motion
selected samples, because of the interlace between the LF and the
velocity distribution function when predicting the behavior of samples
selected {\it simultaneously} in magnitude and proper-motion.

Because the spatial density of white-dwarfs (WDs hereafter) is rather
small (about $3.4 \times 10^{-3}$~stars/pc$^3$ down to $M_V \sim
+16.75$), it is important to insure that the method being used to
determine its LF is either free from biases, or that they can be at
least reliably corrected. Also, it is important to understand the
effects of the kinematic selection on the resulting LF. The purpose of
this research has been precisely the elaboration of simulation tools
that would allow us to answer some of these questions.

Finally, an interesting outcome of all this, is the fact that one
might {\it invert} these simulations and use a well understood
method. In this case, the determined LF can be used in the model to
predict the kinematic distribution of the catalogue stars starting
form a set of assumptions regarding the velocity dispersion of
WDs. The model predictions are then compared to the observations in an
iterative fashion so as to determine the best-fit velocity
distributions that are compatible with the observed LF. The velocity
dispersion of WDs is known only approximately, while its possible
change as a function of luminosity, or cooling age, is completely
unknown (Sion and Liebert 1977) because of the lack of large
homogeneous samples, and the lack of interpretative models. A proper
understanding of the kinematic characteristics of WDs in the solar
neighborhood is critical to an understanding of the dynamical history
of the local disk which is recorded in the WD luminosity
cooling-clock.

\section{Numerical simulations}

The method proposed by Schmidt (1968, 1975) allows for a derivation of
the LF for a sample of stars if we know their apparent magnitudes,
parallaxes and (if used in the sample selection), proper motions. We
also need to know the sample selection (or survey) limits. For more
details of the method, the reader is referred to Schmidt's papers.

Liebert, Dahn \& Monet (1988, LDM88 thereafter) have presented
trigonometric parallaxes, optical colors, and spectrophotometric data
for intrinsically faint WDs in the context of a program to determine
the faint end of the WD LF. Using the classical $1/V_{\rm max}$
method, they derived a LF which indicated a downturn near
$log(L/L_\odot \sim -4.4$, a stellar density of $3 \times
10^{-3}$~stars pc$^{-3}$, and a derived age for the disk in the range
$7-10$~Gyr. More recently, Legget, Ruiz and Bergeron (1998, LRB98
thereafter) gathered new optical and near-IR data for the cool WDs in
the LDM88 sample. Using stellar parameters derived from these data and
more refined model atmospheres they re-derived the faint-end of the WD
LF, also using the $1/V_{\rm max}$ method.  Comparing their LF with
the (then) most recent cooling sequences, they derived a rather young
age for the disk of $8 \pm 1.5$~Gyr. In both cases, the uncertainties
on the LF was computed using the classical approach of assuming
Poisson noise in the counts of every bin, without consideration of the
actual observational errors for the quantities involved in the LF
determination.

For our numerical simulations, it is assumed that observational errors
represent the standard deviation, and that the true value follows a
Gaussian distribution function. In every realization of the LF, a
value with a mean and dispersion from input values is randomly drawn
from a Gaussian distribution function. These values for the whole
sample are then used to construct the LF for that particular
simulation. Collective values, averaged over the simulations, are then
output. In this way, it is possible to derive mean, median \&
quartiles for the LF over a given set of simulations, as well as any
other indicator.

For the simulations, the value adopted for an observable is simply
given by, e.g. for the bolometric correction:

\begin{equation}
bc_{i,j} = bc_j + \sigma_{{\rm bc}_j} \times G_i
\end{equation}

where $bc_j$ is the (mean) observed bolometric correction for star
``j'' in the sample, with ``measurement error'' $\sigma_{{\rm bc}_j}$,
$G_i$ the Gaussian deviate for simulation ``i'', and $bc_{i,j}$ is the
i-th simulation value for the bolometric correction of star j. The
same is performed for proper-motion, apparent magnitude, and parallax
(not distance!).

\section{Results \& Conclusions}
Our simulations indicate that LRB98's data, when properly accounting
for observational errors, does not rule out a disk with an age as
large as 10~Gyr (see Figure~1). This is good news because previous
studies that find ages of 8~Gyr or younger using similar datasets are
difficult to reconcile with an halo age of 15~Gyr (inferred from old
globular clusters) given that Galactic formation and chemical
evolution models suggest a delay of, at most, 3~Gyr between the onset
of star formation in the halo and in the local disk (Wood and Oswalt
1998). Jimenez et al. (1998), using Hipparcos data, have found an
upper limit for the age for the disk field population in the solar
neighborhood of $11 \pm 1$~Gyr, which would be in agreement with our
revised (older) age from the WD LF. Also, we find that current
observational uncertainties and sample sizes do not allow us to
establish the existence of small scale features in the WD LF which
could be indicative of different episodes of star formation in the
disk. This could only be alleviated by dramatically increasing the
currently small samples.

We find, using the $<V/V_{\rm max}>$ completeness criteria, that the
LRB98 sample seems to be missing faint ($M_{bol} > +15.0$), large
proper motion ($\mu > 2.0$~arcsec~yr$^{-1}$) stars, and that the
sample is only complete for $\mu \le 1.5-2.0
$~arcsec~yr$^{-1}$. However, we find that the precise luminosity break
at the faint end of the WD LF is not extremely sensitive to the survey
boundary and/or incompleteness effects.

Finally, we find that most of the current uncertainties in the
observational WD LF come from uncertainties in bolometric corrections
and in parallaxes, while photometry and proper-motions play a minor
role. Therefore, refinements on theoretical models (such that
$\sigma_{\rm BC} \le 0.05$~mag) and parallaxes (with $\sigma_{\pi} \le
1$~mas), as well as larger samples ($N_{\rm samp} \sim 200$, see Wood
and Oswalt 1998), are urgently needed in order to produce a more
refined luminosity function for white dwarfs.

\begin{figure}
\plotfiddle{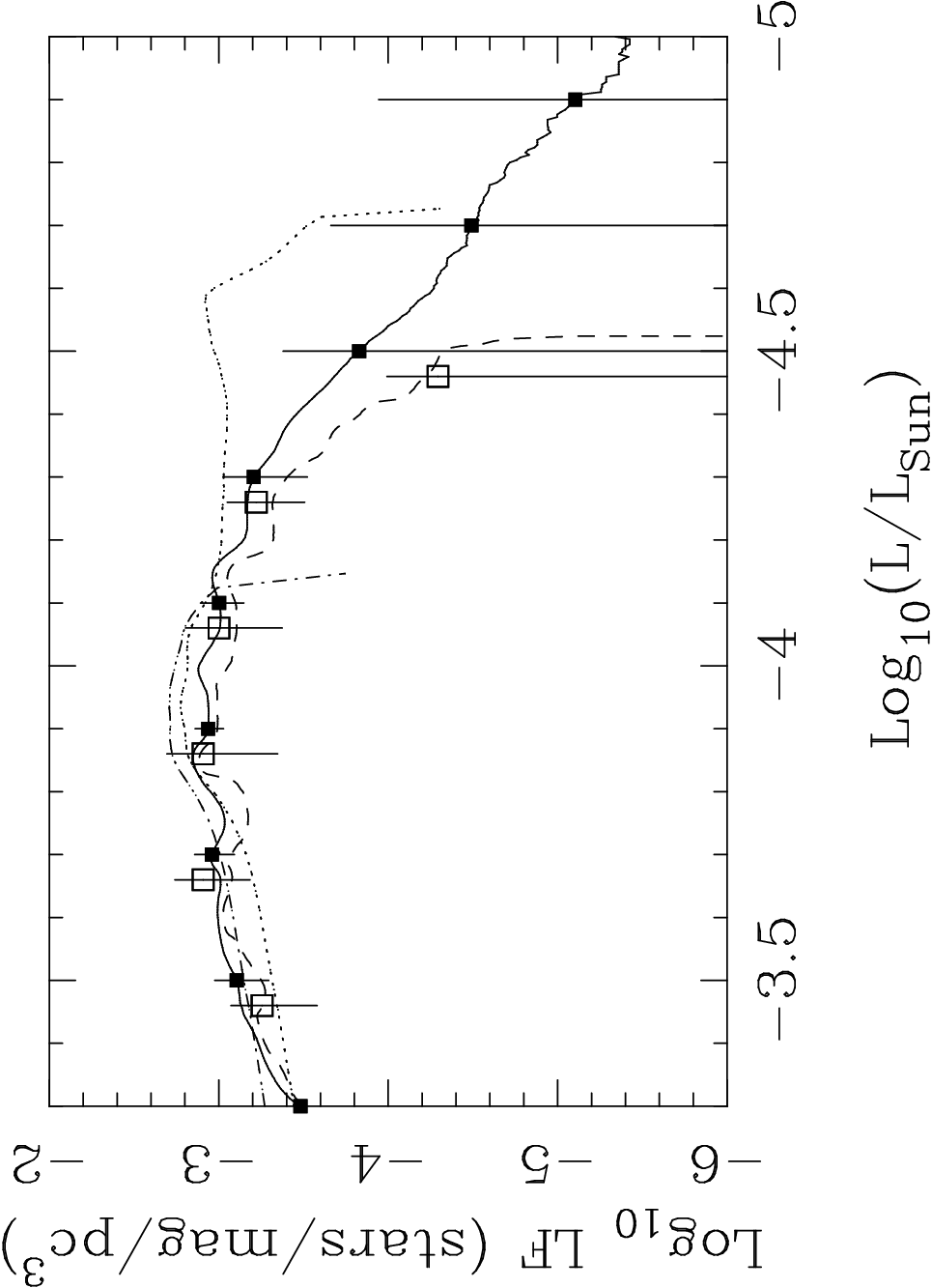}{0.5in}{270}{50}{50}{-190}{290}

\caption{Bolometric LF for the LRB98 dataset. The solid squares with
error bars indicate the LF using the Monte-Carlo mean LF on discrete
0.5~mag bin intervals, while the open squares reproduces the LF
derived in the case of no errors, shifted by +0.04. The solid line
shows the mean LF using a novel moving-box approach. The long-dashed
line indicate the median over simulations LF from the very same
simulation that generated the plotted mean LF. The dot-dashed and
short-dashed lines show the latest theoretical WD~LF published by
Benvenuto and Althaus (1999), based on carbon-oxygen core WDs, for a
6~Gyr and a 10~Gyr disk respectively. The big difference between the
mean and median LF at faint magnitudes indicates a highly skewed
distribution of LF values. Surprisingly, we can also see that the
median LF approaches very well the LF derived in the case of no
errors. We also find that, when using the mean LF, we can not rule
out a disk age of 10~Gyr given the present observational
uncertainties.}

\end{figure}

\end{document}